\documentclass[lineno]{biometrika}

\usepackage{amsmath}
\usepackage{float}

\usepackage{xcolor}
\usepackage{dsfont}
\usepackage{amssymb}
\usepackage{algorithm}
\usepackage[noend]{algpseudocode}
\usepackage{newtxtext}
\usepackage[subscriptcorrection]{newtxmath}
\allowdisplaybreaks 

\newcommand{\E}{E}
\newcommand{\bDelta}{\Delta}
\newcommand{\bdelta}{\delta}
\newcommand{\kn}{K_n}
\newcommand{\Prob}{\text{pr}}
\newcommand{\CE}{\text{CE}}
\newcommand{\N}{\text{N}}
\newcommand{\ind}{\mathds{1}}
\newcommand{\SigmaWtilde}{\Sigma_{\widetilde{W}}}
\newcommand{\SigmaWtildehat}{\widehat{\Sigma}_{\widetilde{W},n}}
\newcommand{\xiCE}{\xi^{\text{CE}}}
\newcommand{\sigmasqWtildehat}{\widehat{\sigma}^2_{\widetilde{W},n}}
\newcommand{\sigmasqWtilde}{\sigma^2_{\widetilde{W}}}
\newcommand{\sigmaWtilde}{\sigma_{\widetilde{W}}}

\makeatletter

\definecolor{editColor}{HTML}{000000}

\begin{document}
\nolinenumbers
\jname{Biometrika}
\jyear{2025}
\jvol{000}
\jnum{0}
\cyear{2025}


\markboth{C. Murphy, E. Laber, R. Merwin, B. Reich, and J. Koerner}{Functional Principal Component Analysis for Sparse Censored Data}
\title{Functional Principal Component Analysis for Sparse Censored Data}

\author{CAITRIN MURPHY and ERIC LABER}
\affil{Department of Statistical Science, Duke University,\\ 214 Old Chemistry, Durham, North Carolina 27708, U.S.A.
\email{caitrin.murphy@duke.edu\quad eric.laber@duke.edu}}

\author{RHONDA MERWIN}
\affil{Department of Psychiatry \& Behavioral Sciences, Duke University,\\ 905 W Main St, Durham, NC 27701
\email{rhonda.merwin@duke.edu}}

\author{BRIAN REICH and JAKE KOERNER}
\affil{Department of Statistics, North Carolina State University,\\ 2311 Stinson Drive, Raleigh, North Carolina 27695, U.S.A.
\email{bjreich@ncsu.edu\quad jrkoerne@ncsu.edu}}

\maketitle

\begin{abstract}
Functional principal component analysis is a key tool in the study of functional data, driving both exploratory analyses and feature construction for use in formal modeling and testing procedures. However, {\color{editColor}existing methods} do not apply when functional observations are censored, e.g., the measurement instrument only supports recordings within a pre-specified interval, thereby truncating values outside of the range to the nearest boundary. A naïve application of existing methods, without correction for instrument-induced censoring, introduces bias into the mean, covariance, and {\color{editColor}functional principal component score} estimators. We extend the {\color{editColor}functional principal component analysis} framework to accommodate noisy, and potentially sparse, censored functional data. Local log-likelihood maximization is used to recover smooth mean and covariance surface estimates that are representative of the latent process's mean and covariance functions. The covariance smoothing procedure yields a positive semi-definite covariance surface, computed without the need to retroactively remove negative eigenvalues in the covariance operator decomposition. Additionally, we construct {\color{editColor}a predictor of the score}, conditional on the censored functional data, and demonstrate its use in the generalized functional linear model. Convergence rates for the proposed estimators are established. In simulation experiments, the proposed method yields better predictive performance and lower bias than existing alternatives. We illustrate its practical value in a study aimed at classifying eating disorder diagnoses in individuals with type 1 diabetes, using censored functional blood glucose data.

\end{abstract}

\begin{keywords}
Functional principal component analysis; Scalar-on-function regression; Censored functional data; Censored predictors.
\end{keywords}

\section{Introduction}\label{sec1}
Functional data are inherently infinite-dimensional, so a finite approximation is required if trajectories are to be used as covariates in a regression, for missing trajectory prediction, or to analyze dominant modes of variation. Functional principal component analysis (FPCA) is a widely used dimension reduction technique that operates under the assumption that functional data are realizations of an $L^2$ process, characterized by continuous mean and covariance functions \citep{Dauxois1982}. Under this assumption, the Karhunen-Loève expansion permits centered functional trajectories to be expressed as infinite sums of eigenfunctions and random {\color{editColor}functional principal component (FPC)} scores (\nocite{Karhunen1947}Karhunen, 1947; \nocite{Loeve1946}Loève, 1946); each random score is a function of a centered realization from the $L^2$ process and an eigenfunction of the covariance surface. The FPCA literature is extensive for functional data observed with or without measurement error, as well as on sparse or dense time grids (e.g.,\nocite{BesseRamsay1986} Besse \& Ramsay, 1986;\nocite{Jamesetal2000} James et al., 2000;\nocite{Yaoetal2005} Yao et al., 2005). Additionally, under the extreme case of fragmentary functional data (also known as functional snippets or incomplete functional data), notable mean and covariance surface estimation techniques have been suggested by \citet{Linetal2021} and \citet{Delaigle2020}.

FPCA techniques are particularly applicable to regression. While mean and covariance functions offer information about the underlying functional process, the fraction of variance explained interpretation of the random scores lends itself to scalar-on-function regression. The first use of the functional linear model is credited to \citet{RamsayDalzell1991}, and key extensions to the generalized functional linear model (GFLM) were made by \citet{MarxEilers1999}, \citet{James2002}, and \citet{MullerStadtmullet2005}. The Karhunen-Loève expansion allows the regression of a scalar-valued response on a functional predictor, with a corresponding $L^2$ coefficient, to be expressed as the regression of the response on an infinite sum of FPC scores and scalar-valued coefficients. Regularization can be achieved by truncating the functional covariate and coefficient FPC expansion to the first $K$ terms that explain a pre-specified fraction of variance. This circumvents the need to add additional penalties, choose knot placement, or spline polynomial degree in advance \citep{Morris2015}.

While the class of techniques for analyzing functional data that are measured directly (or with additional error) is vast, the direct application of existing FPCA and scalar-on-function regression methods to data that are censored is not immediate. Before continuing, it is important to highlight a key distinction in terminology. The word 'censoring' has previously been used to describe both functional data observed on subject-specific time intervals \citep{Delaigle2013}, as well as trajectories subject to right-censoring due to patient dropout or death (Strzalkowska-Kominiak \& Romo, 2021\nocite{Strz2021}; Shi et al., 2021\nocite{Shi2021}). The right-censored setting is particularly relevant to clinical studies, and implies that missing values are concentrated at later timepoints because of incomplete patient follow-up. To address this structure, \cite{Shi2021} proposed a sequential FPCA method that bypasses covariance function estimation and eigendecomposition in favor of a B-spline basis expansion of the mean function and FPCs. The missing values, post-dropout, that are censored on the time domain are presumed to correspond to unobserved values below a threshold. In contrast, we use 'censoring' to refer to instrument-induced censoring of the functional process, meaning the proposed method does not require context-specific censoring assumptions. These censored trajectories are one example of data that are missing not at random. Additionally, we explicitly address censoring at the upper and lower bounds, covariance and eigenfunction estimation for exploratory analyses, and sparse trajectory data.

It may also be natural to refer to such processes as {\em truncated functional data}. However, the word 'truncation' has previously been attached to scalar-on-function regression models for which functional covariates are assumed to influence the response on an unknown subset of the time domain (\nocite{HallHooker2016}Hall \& Hooker, 2016; \nocite{Liuetal2022}Liu et al., 2022).

Lastly, Liu and Houwing-Duistermaat referred to censored trajectories as functional data subject to \textit{detection limits}. They proposed mean (\nocite{LiuHD2022}2022) and covariance (\nocite{LiuHD2023}2023) estimators for the case of one-sided censoring. Fast computation was achieved through a numeric approximation to the likelihood; however, inaccurate estimates of the mean and covariance functions propagate into the FPC scores, so it is not suggested to sacrifice estimation accuracy for computation time. The mean function estimation relies on an estimate of the measurement error variance, which is calculated assuming the censored points are not censored. Additionally, their proposed covariance surface estimate is not necessarily positive semi-definite (PSD) and the oracle bandwidth is used for simulation experiments. The approaches of Liu \& Houwing-Duistermaat (2022; 2023) are further investigated in Section \ref{sec4}, and our procedure appears to achieve lower bias for both the mean and covariance estimation.

Interest in the censored functional data setting is motivated by blood glucose data measured on a continuous glucose monitor (CGM) that only supports readings in the interval $40-400$ mg/dL. The sensor imputes blood glucose levels outside of this range with the nearest boundary, thereby censoring the data. Systematic censoring can also arise in other settings, notably from wearable technology (e.g., a smartwatch) providing continuous physiological monitoring information, as well as from biomarkers with lower detection limits. The proposed procedure is well-suited to longitudinal studies that yield sparse, censored trajectories because the mean and covariance surface estimators are constructed from pooled data.

The naïve application of existing methods to censored functional data induces bias in the mean and covariance surface estimators, which propagates into the random FPC scores. For example, smoothing the sample mean in regions where trajectories are censored from above potentially leads to the under-estimation of the mean function, and the over-estimation of the covariance function on the off-diagonal. A simple option is to treat censored measurements as completely missing data, then take a complete case analysis approach and apply existing methods. However, removing censored points, or censored trajectories entirely, is inefficient and introduces unnecessary sparsity and bias. 

Instead, we propose mean and covariance surface estimation procedures based on local log-likelihood maximization that utilize all of the data to recover the mean and covariance surfaces of the latent process; the censoring status and the behavior in neighboring regions are accounted for simultaneously. Our proposed method is also applicable to non-censored data settings, particularly when the minimum eigenvalue of the sample covariance matrix is near zero. Widespread covariance smoothing techniques pass the sample covariance estimate through a two-dimensional smoother (e.g.,\nocite{Yaoetal2005} Yao et al., 2005;\nocite{RiceSilverman1991} Rice \& Silverman, 1991; \nocite{RamsaySilverman2005}Ramsay \& Silverman, 2005), then remove negative eigenvalues if necessary, which might substantially alter the surface estimate. Instead, we enforce a PSD structure on the covariance surface during the estimation step to avoid this issue. In simulation experiments, the proposed mean and covariance estimation techniques exhibit less bias than current alternatives. Additionally, our suggested extensions to FPCA offer a solution to mitigate the bias incurred by using censored functional data as regressors in a linear model. To the best of our knowledge, the impact of censored regressors in linear models has only been studied for scalar-valued covariates. Notably, \nocite{RigobonStoker2007}\nocite{RigobonStoker2009}Rigobon and Stoker (2007, 2009) extensively described the bias incurred from censored independent variables and showed that the addition of a censoring indicator in a regression model is insufficient. Kong and Nan (2016)\nocite{KongNan2016} proposed a two-stage estimation approach, where the censored tail of the covariate distribution, conditional on the non-censored covariates, is estimated using an accelerated failure time model, then regression parameters are estimated using the pseudolikelihood. We address the functional regressor case, demonstrating the use of our proposed FPC score predictor in the GFLM and deriving convergence rates for the proposed estimators.


\section{Review of Functional Principal Component Analysis}\label{sec2}
\subsection{Functional Principal Component Analysis}
We begin by introducing notation and briefly summarizing the FPCA framework for fully observed functional trajectories, measured with or without error, then we postulate a working GFLM with both functional and scalar-valued covariates. Suppose that there exists an $L^2[0,1]$ process, $Z(\cdot)$, with continuous mean and covariance functions defined by $\mu(t)=\E\{Z(t)\}$ and $\Sigma(s,t)=\text{Cov}\{Z(s),Z(t)\}$. All realizations $Z_i(\cdot)$ are assumed to be independent for $i=1,\dots,n$. Application of Mercer's theorem yields the covariance operator decomposition $\Sigma(s,t)=\sum\limits_{k\geq 1}\lambda_k\phi_k(s)\phi_k(t)$ $\forall s,t\in [0,1]$, where $\{\lambda_k, \phi_k(\cdot)\}_{k\geq 1}$ are the eigenpairs. Eigenvalues are ordered so that $\lambda_1\geq\lambda_2\geq\dots\geq 0$ and satisfy $\sum_{k=1}^\infty \lambda_k < \infty$; the eigenfunctions are orthonormal, i.e., $\int_0^1 \phi_k(t)\phi_j(t)dt=1$ for $k=j$, and $0$ otherwise. The $k$th FPC score for unit $i$ is defined by 
\begin{align}
    \xi_{i,k}=\int_0^1\{Z_i(t)-\mu(t)\}\phi_k(t)dt,\label{fpcsocre}
\end{align}
and the Karhunen-Loève expansion allows each functional trajectory to be expressed as, $Z_i(t)=\mu(t)+\sum\limits_{k\geq 1} \xi_{i,k}\phi_k(t)$, $\forall t\in[0,1]$.

In the presence of measurement error, the noisy surrogate $\widetilde{W}_i(\cdot) = Z_i(\cdot)+\sigma(\cdot)\epsilon_i(\cdot)$ is observed for each unit $i$, where $\epsilon_i(\cdot)$ is a Gaussian white noise process, independent across subjects, with covariance operator $\Sigma_{\epsilon}(s,t)=1$ if $s=t$ and 0 otherwise. The measurement error variance, $\sigma^2(\cdot)$, is assumed to be in $L^2[0,1]$. Suppose each $\widetilde{W}_i(\cdot)$ is recorded at ordered times $T_i=(T_{i,1},\dots,T_{{i,N_i}})$; the number of recordings, $N_i$, and all elements in $T_i$ are random variables, thereby allowing for sparse and irregularly observed data. The number of measurements is assumed to be independent of observation times and measurements for each unit $i$, and the observation times are assumed to be iid; both assumptions are consistent with \citet{Yaoetal2005}. Noisy measurements from the latent process observed at times $T_i$ are denoted by $\widetilde{W}_i(T_i) = \{\widetilde{W}_i(T_{i,1}), \ldots, \widetilde{W}_i(T_{i,N_i})\}$. Similarly, we denote functions $\mu(\cdot)$, $\sigma^2(\cdot)$, and $\phi_k(\cdot)$ evaluated at $T_i$ by $\mu(T_i), \sigma^2(T_i), $ and $\phi_k(T_i)\in\mathbb{R}^{N_i}$, respectively. Define the $N_i\times N_i$ matrix $\SigmaWtilde (T_i,T_i) = \Sigma(T_i,T_i) + \text{diag}\{\sigma^2(T_i)\}$, with $(j,k)$th element equal to $\SigmaWtilde (T_{i,j}, T_{i,k})= \Sigma(T_{i,j}, T_{i,k}) + \sigma^2(T_{i,j})\ind_{j=k}$. Estimating the FPC scores from the noisy data is not immediate because a direct replacement of $Z_i(\cdot)$ with $\widetilde{W}_i(\cdot)$ in (\ref{fpcsocre}) will be biased. The seminal solution, proposed by \citet{Yaoetal2005}, uses the assumed joint normality of the random scores and noisy data to construct the best linear unbiased predictor of the $k$th FPC score for unit $i$. The predictor for the $k$th FPC score for unit $i$ is defined by the conditional expectation, $\xiCE_{i,k} = \E\{\xi_{i,k}\mid \widetilde{W}_i(T_i)=\widetilde{w}_i(t_i)\} = \lambda_k \phi_{k}(t_i)^\intercal\SigmaWtilde (t_i,t_i)^{-1}\{\widetilde{w}_i(t_i)-\mu(t_i)\}$. An estimator of $\xiCE_{i,k}$ is $\widehat{\xi}^{\CE}_{i,k,n}= \widehat{\lambda}_{k,n} \widehat{\phi}_{k,n}(t_i)^\intercal\SigmaWtildehat(t_i,t_i)^{-1}\{\widetilde{w}_i(t_i)-\widehat{\mu}_n(t_i)\}$, where $\widehat{\mu}_n(t_i)$, $\SigmaWtildehat(t_i,t_i)$, and $\{\widehat{\lambda}_{k,n},\widehat{\phi}_{k,n}(t_i)\}$ are estimators of $\mu(t_i)$, $\SigmaWtilde (t_i,t_i)$, and the $k$th eigenpair $\{\lambda_k,\phi_k(t_i)\}$, respectively. Eigenpair estimates are obtained from an approximate discretization of the functional eigenequation for $\Sigma(\cdot,\cdot)$ \citep{RamsaySilverman2005}. Next, we discuss the use of the FPC score predictors in the GFLM.

\subsection{Scalar-on-Function Regression}
In addition to the noisy functional measurements, suppose that $p$ baseline covariates, $X_i\in\mathbb{R}^p$, and an outcome, $Y_i\in\mathcal{Y}\subseteq \mathbb{R}$, are recorded for each unit. The augmented data are of the form $[\{\widetilde{W}_i(T_i), X_i, Y_i\}]_{i=1}^n$ and the $n$ copies are assumed to be iid. We assume that the conditional mean of the outcome given the latent functional process, $Z(\cdot)$, and the covariates, $X$, is
\begin{align*}
g \left[\E \left\{Y_i\mid X_i=x_i, Z_i(\cdot) = z_i(\cdot) \right\} \right] = & \alpha_0+x_i^\intercal\alpha +\int_0^1z_i(t)\beta(t)dt\\
=&\alpha_0+x_i^\intercal\alpha +\int_0^1\mu(t)\beta(t)dt+\sum\limits_{k\geq 1}\beta_k\xi_{i,k},
\end{align*}
where $g(\cdot)$ is a link function, $\alpha_0\in\mathbb{R}$ is the intercept, and $\alpha\in\mathbb{R}^p$ is the vector of coefficients corresponding to the baseline covariates. The coefficient function $\beta(\cdot)$ weights the functional predictor across time and is assumed to be in $L^2[0,1]$; the second equality follows from the expansion $\beta(t)=\sum\limits_{k\geq 1}\beta_k\phi_k(t)$. As $Z(\cdot)$ is not directly observable, we reformulate the model in terms of the available data
\begin{align}
    &g \left[\E \left\{Y_i\mid X_i=x_i, \widetilde{W}_i(T_i)=\widetilde{w}_i(t_i) \right\} \right]\nonumber\\
    &=g\Big[\E \Big\{g^{-1}\Big(\alpha_0^*+x_i^\intercal\alpha +\sum\limits_{k\geq 1}\beta_k\xi_{i,k} \Big)\mid X_i=x_i, \widetilde{W}_i(T_i)=\widetilde{w}_i(t_i) \Big\} \Big], \nonumber
\end{align}
which follows from the assumptions that (A1) $Y_i\perp \widetilde{W}_i(\cdot)\mid \{Z_i(\cdot), X_i\}$, meaning the noisy trajectory offers no additional information to the outcome, given the true trajectory and baseline covariates, and (A2) FPC scores are independent across trajectories and uncorrelated across $k$, each $\xi_{i,k}\sim\mathcal{N}(0, \lambda_k)$, $\lambda_1\geq\lambda_2\geq\dots\geq 0$, and $\sum_{k=1}^\infty \lambda_k < \infty$. We define $\alpha_0^*=\alpha_0 + \int_0^1\mu(t)\beta(t)dt$. The assumption of joint normality of the scores and $\widetilde{W}_i(\cdot)$ permits the direct evaluation of the expectation. In the following section, we extend the FPCA framework to accommodate censored functional data, and illustrate the use of the FPC score predictors for modeling.

\section{FPCA for Censored Data}\label{sec3}
Instead of directly observing either $Z_i(\cdot)$ or $\widetilde{W}_i(\cdot)$, suppose we record a noisy surrogate censored to the interval $[a,b]$, denoted by $W_i(\cdot)$. The censored trajectory is $W_i(\cdot)=\text{max} [a, \text{min} \{b,\widetilde{W}_i(\cdot) \} ]$. Define the censoring indicator matrix for unit $i$ as $\Delta_i(T_i)$; rows are of the form $\Delta_i(T_{i,j})=(\delta^a_{i,j},\delta^0_{i,j}, \delta^b_{i,j})$, where $\delta_{i,j}^{a}=\delta_{i}^{a}(T_{i,j})=\ind\{\widetilde{W}_i(T_{i,j})\leq a\}$, $\delta_{i,j}^{0}=\delta_{i}^{0}(T_{i,j})=\ind\{\widetilde{W}_i(T_{i,j})\in (a,b)\}$, and $\delta_{i,j}^{b}=\delta_{i}^{b}(T_{i,j})=\ind\{\widetilde{W}_i(T_{i,j})\geq b\}$ represent whether an observation at $T_{i,j}$ is censored from below, not censored, or censored from above. Consequently, the observed data are $\mathcal{D}_n=\Big[\{W_i(T_i), \Delta_i(T_i), X_i, Y_i\}\Big]_{i=1}^n$.

Prior to estimating the best linear unbiased predictors for the FPC scores, the mean and covariance functions must be estimated. Often, non-parametric methods are used to smooth the sample mean and covariance estimators, however, using the same techniques on the censored data introduces additional bias in the surface estimates. Our method centers on constructing estimators of $\mu(\cdot)$ and $\Sigma(\cdot,\cdot)$ that account for censoring through the maximization of the relevant local kernel-weighted log-likelihood functions. We obtain the FPC scores through a three-stage procedure: 1) construct estimators $\widehat{\mu}_n(t)$ of $\mu(t)$ and $\sigmasqWtildehat(t)$ of $\sigmasqWtilde(t)$ over a pre-specified grid, 2) construct estimators $\SigmaWtildehat(s,t)$ of $\SigmaWtilde (s,t)$ and $\widehat{\Sigma}_n(s,t)$ of $\Sigma(s,t)$ over the same grid as in 1), then 3) under the assumption that the random scores and latent noisy process $\widetilde{W}(\cdot)$ are jointly normal, formulate the FPC score predictors.

Beginning with the first stage, let $A_h: \mathbb{R}\rightarrow \mathbb{R}$ denote a kernel function with bandwidth $h > 0$; in our applications we use a Gaussian kernel, but this is not essential. Define the observed-data local log-likelihood \citep{TibshiraniHastie1987} at point $t\in[0, 1]$ as,
\begin{align*}
\ell_{h,n}\{&t; \mu(t), \sigmasqWtilde(t)\} = \sum\limits_{i=1}^n\sum\limits_{j=1}^{N_i} A_h(t-t_{i,j})\Big[\delta_{i,j}^{a}\log\left\{ \Phi\left(\tfrac{a-\mu(t)}{\sigmaWtilde(t)}\right)\right\} \\
& + \delta_{i,j}^0\log\left[\frac{1}{\sigmaWtilde(t)}\phi\left\{\tfrac{W_i(t_{i,j})-\mu(t)}{\sigmaWtilde(t)}\right\}\right]+\delta_{i,j}^{b}\log\left\{ 1-\Phi\left(\tfrac{b-\mu(t)}{\sigmaWtilde(t)}\right) \right\}\Big].
\end{align*}
Maximizing $\ell_{h,n}\{t;\mu(t),\sigmasqWtilde(t)\}$ at $t\in[0,1]$ yields the local estimators $\widehat{\mu}_n(t)$ and $\sigmasqWtildehat(t)$. The bandwidth is tuned using leave-one-curve-out cross-validation \citep{RiceSilverman1991}. We note that the contributions to $\ell_{h,n}\{t;\mu(t),\sigmasqWtilde(t)\}$ from censored points are non-positive, and potentially disproportionately smaller than contributions from non-censored points in the same region. Simulation results suggest that in such cases, the cross-validation procedure attempts to include more non-censored points than what is optimal, thereby favoring larger bandwidths. We suggest that cross-validation should be performed exclusively on the non-censored observed points. The estimated optimal bandwidth is chosen to minimize the criterion function $\text{CV}_n(h)=\sum_{i=1}^n \sum_{j=1}^{N_i} \delta_{i,j}^0\{W_i(t_{i,j}) - \widehat{\mu}_{n}^{0,(-i)}(t,h)\}^2$, where $\widehat{\mu}_{n}^{0,(-i)}(t,h)$ is the local mean estimate at time $t$ under bandwidth $h$, computed from the data set excluding the $i$th trajectory and all censored observations. Following bandwidth selection, local estimates of the mean function and variance-plus-measurement-error function are $\{\widehat{\mu}_n(t), \sigmasqWtildehat(t)\}=\underset{\mu, \widetilde{\sigma}^2}{\text{argmax }}\ell_{h,n}\{t;\mu(t),\sigmasqWtilde(t)\}$. Measurement error variance and diagonal covariance elements are separated in the next step.

For the covariance surface estimation step, we begin by iteratively maximizing the local log-likelihood for each off-diagonal covariance element, under the constraint that each element update maintains the PSD structure of $\SigmaWtilde (\cdot,\cdot)$. Denote the set of upper triangular covariance matrix indices for unit $i$ as $\mathcal{J}_i = \{(j, j') : j=1,\dots,N_i, j' = 1, \dots, N_i, j < j'\}$. Let $B_{(h_1,h_2)}: \mathbb{R}\times \mathbb{R}\rightarrow \mathbb{R}$ denote a two-dimensional kernel function with bandwidths $h_1,h_2 > 0$. Because $\Sigma(\cdot,\cdot)$ is a symmetric operator, we choose $h=h_1=h_2$; as a result, the notation $B_{(h_1,h_2)}$ simplifies to $B_h$. In our implementation we used a bivariate Gaussian kernel. The local log-likelihood for the off-diagonal element $\sigma(s,t)$ is,
\begin{align*}
    \ell_{h,n}&\{s,t;\sigma(s,t),\mu(s), \mu(t),\sigmasqWtilde(s), \sigmasqWtilde(t) \} \nonumber\\ = &\sum_{i=1}^n\sum_{(j,j')\in\mathcal{J}_i}
    B_{h}(s - t_{i,j}, t - t_{i,j'})\nonumber\\
    &\times\Big({\delta_{i,j}^0\delta_{i,j'}^0}\log[ p\{W_i(t_{i,j}), W_i(t_{i,j'})\}]
    \\
    &+\text{ }\text{ }{\delta_{i,j}^a\delta_{i,j'}^a}
    \log[ \Prob\{W_i(t_{i,j})\leq a, W_i(t_{i,j'})\leq a\}]
    \\
    &+\text{ }\text{ }{\delta_{i,j}^b\delta_{i,j'}^b}\log[ \Prob\{W_i(t_{i,j})\geq b, W_i(t_{i,j'})\geq b)]
    \\
    &+\text{ }\text{ }{\delta_{i,j}^a\delta_{i,j'}^b}\log[ \Prob\{W_i(t_{i,j})\leq a, W_i(t_{i,j'})\geq b\}]
    \\
    &+\text{ }\text{ }{\delta_{i,j}^b\delta_{i,j'}^a}\log[ \Prob\{W_i(t_{i,j})\geq b, W_i(t_{i,j'})\leq a\}]
    \\
    &+\text{ }\text{ }{\delta_{i,j}^a\delta_{i,j'}^0}\log\left[ \Prob\{W_i(t_{i,j})\leq a \mid W_i(t_{i,j'})\in(a,b)\} \times \phi\left\{\tfrac{W_i(t_{i,j'}) - \mu(t)}{\sigmaWtilde(t)}\right\}\right]
    \\
    &+\text{ }\text{ }{\delta_{i,j}^b\delta_{i,j'}^0}\log\left[ \Prob\{W_i(t_{i,j})\geq b \mid W_i(t_{i,j'})\in(a,b)\} \times \phi\left\{\tfrac{W_i(t_{i,j'}) - \mu(t)}{\sigmaWtilde(t)}\right\}\right]
    \\
    &+\text{ }\text{ }{\delta_{i,j}^0\delta_{i,j'}^a}\log\left[ \Prob\{W_i(t_{i,j'})\leq a \mid W_i(t_{i,j})\in(a,b)\} \times \phi\left\{\tfrac{W_i(t_{i,j}) - \mu(s)}{\sigmaWtilde(s)}\right\}\right]
    \\
    &+\text{ }\text{ }{\delta_{i,j}^0\delta_{i,j'}^b}\log\left[ \Prob\{W_i(t_{i,j'})\geq b \mid W_i(t_{i,j})\in(a,b)\} \times \phi\left\{\tfrac{W_i(t_{i,j}) - \mu(s)}{\sigmaWtilde(s)}\right\}\right]\Big),
\end{align*}
\noindent
{\color{editColor} where $p$ is a bivariate Gaussian density with mean $\mu_{s,t} = \{\mu(s), \mu(t)\}^\intercal$ and covariance matrix $\Psi_{s,t}$, which is composed of off-diagonal element $\sigma(s,t)$ and variance elements $\sigmasqWtilde(s)$ and $\sigmasqWtilde(t)$; the unconditional probabilities are computed under the corresponding cdf; the conditional probabilities rely on conditional mean and variance parameters that follow from results on partitioned multivariate Gaussian vectors (see the Supplementary Material for further details).}

The off-diagonal elements of both $\widehat{\Sigma}_n(\cdot,\cdot)$ and $\SigmaWtildehat(\cdot,\cdot)$, are populated with $\widehat{\sigma}_{n}(s,t) = \underset{\sigma(s,t)}{\text{argmax }}\ell_{h,n}\{s,t;\sigma(s,t),\widehat{\mu}_n(s), \widehat{\mu}_n(t),\sigmasqWtildehat(s), \sigmasqWtildehat(t)\}$. Estimators of $\mu_{s,t}$ and $\{\sigmasqWtilde(s), \sigmasqWtilde(t)\}$ are available from the previous stage. Projected gradient descent is used to maintain the PSD structure of $\SigmaWtilde (\cdot,\cdot)$ during optimization. Let $\SigmaWtildehat(\cdot,\cdot)$ denote the matrix with $(s,t)$th element $\widehat{\sigma}_n(s,t)\ind_{s\ne t} + \sigmasqWtildehat(s)\ind_{s=t}$. For notational convenience, define the set of timestamps for non-censored recordings from the $i$th trajectory by $T_i^0 = \{T_{i,j}:\delta_{i,j}^0=1\}$. The corresponding vector of non-censored measurements is $W_i(T_i^0)$, and the matrix $\SigmaWtildehat(T_i^0,T_i^0)$ has the analogous rows and columns of $\SigmaWtildehat(\cdot,\cdot)$. While leave-one-curve-out cross-validation may be used to select the bandwidth, a less computationally intensive option is to select the bandwidth $h$ (or the bandwidth pair $(h_s,h_t)$) that maximizes the pseudo-likelihood $\widehat{\mathcal{L}}_n=\prod_{i=1}^n p_0\{W_i(T^0_i)\}$, where $p_0$ denotes the Gaussian density with mean $\widehat{\mu}_n(T_i^0)$ and covariance $\SigmaWtildehat(T_i^0,T_i^0)$ for the non-censored data from unit $i$.

Assuming the measurement errors and stochastic realizations are independent, a smooth estimator of $\Sigma(\cdot,\cdot)$ can be obtained by removing the diagonal elements of $\SigmaWtildehat(\cdot,\cdot)$, then applying a two-dimensional local smoother to estimate the diagonal elements of $\Sigma(\cdot,\cdot)$. We follow \nocite{Yaoetal2003}Yao et al.'s (2003) suggestion to use a local quadratic term for the direction perpendicular to the diagonal, and local linear term for the direction parallel to the diagonal. Define the local smoother $D_{h,n}\{s,t;\gamma,\SigmaWtildehat(T,T)\}=\sum\limits_{i=1}^n\sum\limits_{1\leq j\ne j'\leq N_i} B_h(s-T_{i,j},t-T_{i,j'})[\SigmaWtildehat(T^*_{i,j},T^*_{i,j'})-\{\gamma_0+\gamma_1(s-T_{i,j}^*)+\gamma_2(t-T^*_{i,j'})^2\}]$, where $B_h(\cdot,\cdot)$ is a two-dimensional Gaussian kernel, and the time index pair $(T^*_{i,j},T^*_{i,j'})$ represents the $\pi/4$ radian rotation of $(T_{i,j},T_{i,j'})$ to the right \citep{Yaoetal2003}. The rotation of the axes permits the direct use of $h$ from the $\sigmasqWtilde(t)$ optimization step, negating the formulation of an additional bandwidth selection criterion. However, the use of $D_{h,n}$ does not guarantee a resulting PSD matrix, so we enforce the additional constraints that 1) $\widehat{\gamma}_{0,n}(t) \leq \sigmasqWtildehat(t)$ and 2) the singular value decomposition of $\widehat{\Sigma}_n(T,T)$ yields non-negative eigenvalues using projected gradient descent. Under these constraints, the smoothed diagonal estimators of the covariance operator are $\widehat{\Sigma}_n(t,t)=\widehat{\gamma}_{0,n}(t)$, from $\underset{\gamma}{\text{argmin }}D_{h,n}\{s,t;\gamma,\SigmaWtildehat(T,T)\}$. Consequently, the measurement error variance estimator is $\widehat{\sigma}^2_n(t)=\SigmaWtildehat(t,t) - \widehat{\Sigma}_n(t,t)$. Eigenvalue and eigenvectors of the form $\{\lambda_k,\phi_k(T)\}$ may be estimated for $1\leq k\leq N$ from the $N\times N$ matrix $\widehat{\Sigma}_n(T,T)$. Further details regarding the algorithm used for the covariance optimization are provided in the Supplementary Material. 

To construct the random score predictors for the third stage, we adapt the conditional expectation approach \citep{Yaoetal2005}. Recall that the non-censored data and FPC scores are jointly normal, so we may first consider treating the censored points as completely missing, thereby ignoring the censoring indicator matrix. The best linear unbiased predictor of the $k$th FPC for unit $i$ follows immediately: $\xi^{\CE,0}_{i,k}=\E\{\xi_{i,k}\mid  W_i(T_i^0)=w_i(t^0_i)\}=\lambda_k \phi_k(t^0_i)^\intercal \SigmaWtilde (t_i^0, t_i^0)^{-1}$ $\{w_i(t_i^0)-\mu(t_i^0)\}$. However, much of the data are potentially removed if trajectories have large censored regions. To avoid a substantial discard of data, we instead construct a best linear unbiased predictor, conditional on all of the observed trajectory data and censoring indicators: $\xi^*_{i,k}=\E\{\xi_{i,k}\mid  W_i(T_i)=w_i(t_i), \Delta_i(T_i)=\delta_i(t_i)\}$. The predictor can be rewritten as,
\begin{align}
    \xi^*_{i,k} =& \E\{\xi_{i,k}\mid W_i(T_i)=w_i(t_i), \Delta_i(T_i)=\delta_i(t_i)\}\nonumber\\
    =& \E[\E\{\xi_{i,k}\mid \widetilde{W}_i(T_i)= \widetilde{w}_i(t_i)\}\mid W_i(T_i)=w_i(t_i),\Delta_i(T_i)=\delta_i(t_i)]\label{MCind}\\
    =& \E\{\xiCE_{i,k}\mid W_i(T_i)=w_i(t_i),\Delta_i(T_i)=\delta_i(t_i)\}\nonumber,
\end{align}
where ($\ref{MCind}$) follows from the assumption (A3) $Z_i(\cdot)\perp \{W_i(\cdot),\Delta_i(\cdot)\}\mid \widetilde{W}_i(\cdot)$, and $\xiCE_{i,k}$ is the best linear unbiased predictor described in Section \ref{sec2}. However, $\xiCE_{i,k}$ is not directly estimable because $\widetilde{W}_i(T_i)$ is not fully observed in censored regions. Therefore, we approximate $\xi_{i,k}^*$ with the estimator $\widehat{\xi}_{i,k,n}$, which is computed by sampling $m$ vectors, $\widetilde{W}_i(T_i)\mid W_i(T_i)=w_i(t_i),\Delta_i(T_i)=\delta_i(t_i)$, then plugging each vector into $\widehat{\xiCE}_{i,k,n}$ from Section \ref{sec2} and taking the average over the $m$ estimates.

Sampling the latent noisy discretized process, conditional on the observed data and censoring indicators, is equivalent to sampling the censored regions, conditional on the non-censored data and all of the censoring indicators, and updating $\widetilde{W}_i(t_i)$ accordingly. To do so, reorder $t_i$ for each unit $i$ such that the non-censored data indices are first and the censored data indices are grouped together after; this reordered time index set is denoted by $\widetilde{t_i}$. The partially observed noisy process for unit $i$ on the observed grid $\widetilde{t_i}$ may be partitioned as follows: $\widetilde{W}_i(\widetilde{t}_i)=\{\widetilde{W}_i(\widetilde{t}^0_i), \widetilde{W}_i(\widetilde{t}_i\setminus \widetilde{t}^0_i) \}^\intercal=\{\widetilde{W}_{i,1}, \widetilde{W}_{i,2} \}^\intercal$ with $\widetilde{W}(\widetilde{t}_i)\sim \mathcal{N}\{\mu(\widetilde{t}_i), \SigmaWtilde (\widetilde{t}_i,\widetilde{t}_i)\}$. Previously estimated mean and covariance parameters are plugged in for $\mu(\widetilde{t}_i)$ and $\SigmaWtilde (\widetilde{t}_i,\widetilde{t}_i)$. Results on partitioned multivariate Gaussian vectors are utilized to sample $m$, $|T_i\setminus T_i^0|$-dimensional vectors from $\widetilde{W}_{i,2}\mid W_i(T_i)=w_i(t_i), \bDelta(T_i)=\bdelta(t_i) \sim \text{Truncated Normal}(\mu_{\widetilde{W}_{i,2}\mid \widetilde{W}_{i,1}},\Sigma_{\widetilde{W}_{i,2} \mid \widetilde{W}_{i,1}}, a^*, b^*)$, where $a^*$ has elements set to $b$ when $\delta^b_i(t)=1$ (i.e., $\ind\{\widetilde{w}_{2,i}\geq b \}=1$) and $-\infty$ when $\delta^b_i(t)=0$ (not censored or censored at the lower bound) and $b^*$ has elements set to $a$ when $\delta_i^a(t)=1$ (i.e., $\ind\{\widetilde{w}_{2,i}\leq a \}=1$) and $\infty$ when $\delta^a(t)=0$ (not censored or censored at the upper bound). Subsequent samples are populated in the relevant entries of $\widetilde{W}_i(t_i)$, then the $m$ $\widehat{\xi}^{\CE}_{i,k,n}$'s are computed and averaged together, yielding $\widehat{\xi}_{i,k,n}$. Under the assumption that the $i$th functional process may be well-approximated by the first $K$ FPC scores, the predicted trajectory for unit $i$ is $\widehat{W}^K_{i,n}(t)=\widehat{\mu}_n(t)+\sum\limits_{k=1}^K\widehat{\xi}_{i,k,n}\widehat{\phi}_{k,n}(t)$.

Now, we illustrate the use of $\widehat{\xi}_{i,k,n}$ in the GFLM,
\begin{align}
    g [\E \{ &Y_i\mid X_i=x_i, W_i(T_i)=w_i(t_i), \Delta_i(T_i)=\delta_i(t_i) \} ] \nonumber\\
    &=g\Big(\E \Big[\E \Big\{g^{-1}\Big(\alpha_0^*+x_i^\intercal\alpha +\sum_{k\geq 1}\beta_k\xi_{i,k} \Big)\mid X_i=x_i, \widetilde{W}_i(T_i)=\widetilde{w}_i(t_i) \Big\}\Big| \\
    &\qquad  X_i = x_i, W_i(T_i)=w_i(t_i), \bDelta_i(T_i)=\bdelta(t_i) \Big]\Big) \nonumber,
\end{align}
where the equality follows from assumptions (A4) $Y_i\perp \{W_i(\cdot), \Delta_i(\cdot)\} \mid\{Z_i(\cdot),X_i\}$ and (A5) $Y_i\perp \{W_i(\cdot), \Delta_i(\cdot)\} \mid\{\widetilde{W}_i(\cdot),X_i\}$. Assumption (A4) implies that trajectory noise and censoring information do not provide additional information to unit $i$'s outcome, given the true functional realization and baseline covariates, and (A5) states that unit $i$'s outcome is independent of censoring information, given knowledge of the noisy non-censored process and baseline covariates. We assume the regression is reasonably approximated by the first $K$ FPC scores, where $K$ is selected by the fraction of variance explained. The fraction of variance explained by the $K$th FPC score is defined as $\sum_{j=1}^K\lambda_j / \sum_{j=1}^\infty \lambda_j$. The coefficients $(\alpha_0^*, \alpha^{\intercal}, \beta_1,\dots,\beta_K)$ may be estimated via maximum likelihood.

We establish convergence rates for the GFLM when $Y_i\in\mathbb{R}$ and $g(\cdot)$ is the identity link. Let $M\{x, w(t),\bdelta(t)\}=\E\{Y\mid X=x, W(T)=w(t), \bDelta(T)=\bdelta(t)\}$. In addition to (A1)-(A5), the following assumptions are used: (A6) $Z_i(\cdot)\perp X_i \mid \widetilde{W}_i(\cdot)$, implying that the true process offers no additional information about the baseline covariates, given the noisy trajectory; (A7) $\lim\limits_{n\to \infty}\bigcup\limits_{i=1}^nT_i$ forms a dense set in $[0,1]$ with probability 1; (A8) $\underset{t}{\sup}|\widehat{\mu}_n(t)-\mu(t)|=O_p(n^{-\zeta_{\mu})}$ for $t\in[0,1]$ and some positive $\zeta_{\mu}$; (A9) $\underset{s,t}{\sup }|\widehat{\Sigma}_n(s,t)-\Sigma(s,t)|=O_p(n^{-\zeta_{\Sigma}})$ for $s,t\in[0,1]$ and some positive $\zeta_{\Sigma}$; (A10) $\underset{t }{\sup }|\widehat{\sigma}_n^2(t)-\sigma^2(t)|=O_p(n^{-\zeta_{\sigma}})$ for $t\in[0,1]$ and some positive $\zeta_{\sigma}$; (A11) $\E|| X|| ^ 4$ and $\E(XX^\intercal)$ is nonsingular; (A12) $|\beta_k|\leq Ck^{-\eta}$ for all $k\geq 1$, and some $\eta >1$ and $C > 0$; (A13) $N\leq q$ for some $q < \infty$ with probability 1; and (A14) $\exists r, 0\leq r<1$ such that $\sum_{i=1}^n\sum_{j=1}^{N_i}(\delta^a_{i,j}+\delta^b_{i,j}) / (\sum_{i=1}^nN_i) \leq r$ as $n\to \infty$. Assumptions (A8)-(A10) are consistent with the sparse functional data analysis literature (Laber \& Staicu, 2018; Yao et al., 2005; Li \& Hsing, 2010\nocite{LiHsing2010}).
\begin{theorem}
Let $K_n$ be an increasing sequence of integers such that $K_n\to\infty$ and $\kn/n^{2\zeta}\to 0$ as $n\to\infty$. Under assumptions (A1)-(A14),
$$\E\Big[\Big|\widehat{M}_n^{\kn}\{X,W(T),\bDelta(T);\widehat{\theta}^{\kn}_n\} - M\{X,W(T),\bDelta(T)\} | \mathcal{D}_n\Big|\Big]=O_p(\kn n^{-1/2}+\kn^{1/2}n^{-\zeta}),$$
where $\zeta = \min(\zeta_{\mu},\zeta_{\Sigma},\zeta_{\sigma})$ and $\widehat{\theta}^{\kn}=(\widehat{\alpha}^*_{0,n},\widehat{\alpha}^\intercal_{n}, \widehat{\beta}_{1,n},\dots,\widehat{\beta}_{\kn,n} )^\intercal.$\nocite{LaberStaicu2018}
\end{theorem}
\noindent
The proof is provided in the Supplementary Material.

\section{Simulation Study}\label{sec4}
We evaluate the finite sample performance of the proposed estimator in a suite of simulation experiments. As a baseline, we also implemented the proposed estimation procedure under the incorrect assumption that the data are free from censoring (termed the naïve method) and the canonical PACE method \citep{Yaoetal2005}, which also does not account for censoring. We also carried out the complete case analysis approach, dropping all censored points from each trajectory (simulation results are provided in the Supplementary Material); the results showed poor performance in comparison to the proposed method. Performance is measured in terms of the SSE for the mean, covariance, measurement error variance, and signal-to-noise ratio (SNR) estimators; for the regression settings, either the mean square error and adjusted $R^2$ or classification accuracy and AIC are used, depending on the model. For all simulations: $t\in[0,1]$, the true mean function is $\mu(t) = \text{sin}(2\pi t)$, and the measurement error is chosen so that the SNR resembles that from the CGM device discussed in Section \ref{sec5} \citep{Nagletal2021}. The observation interval is discretized to a grid of $g=15$ equally spaced points, $t=\{t_1,\dots,t_g\}$, and for each unit, the number of observations is sampled from $N_i\sim \text{Uniform}\{5,\dots,g\}$, then timestamps are sequentially sampled from $T^{(1)}\mid N_i = n_i\sim \text{Uniform}\{t\}$ and $T^{(j+1)}\mid N_i = n_i, T^{(1)}=t^{(1)},\dots, T^{(j)}=t^{(j)}\sim \text{Uniform}[t \setminus \{\bigcup\limits_{k=1}^jt^{(k)}\} ]$. The subsequent ordered observation times for unit $i$ are $T_i =\{ \bigcup_{j=1}^{N_i}t^{(j)}\}$. 

Ten simulation settings are considered. The choices of covariance operator and measurement error variance differentiate the cases. The first four cases illustrate the performance of the mean and covariance estimation procedures as the covariance structure builds in complexity. The remaining six cases demonstrate the recovery of the latent trajectory, $Z(\cdot)$, and the FPC scores, as well as the use of the FPC score estimator in the GFLM. The fraction of variance explained distinguishes \textit{Case 5} and \textit{Case 6}. To numerically assess the sensitivity of the estimators to normality assumptions, non-Gaussian generative models for $\xi_{i,k}$ and $\epsilon_i(\cdot)$ are considered in \textit{Cases 7} - \textit{10}.

For the first four cases, discretized measurements from $n=100$ trajectories are generated from the model $W_i\mid N_i=n_i, T_i=t_i,\overset{\text{iid}}{\sim} \mathcal{N}\{\mu(t_i), \SigmaWtilde (t_i,t_i)\}$, then censored so that all $w_i(t_{i,j})\in[-1,1]$. For \textit{Cases 5} - \textit{10}, the $n=100$ discretized trajectories are constructed from $W_i = \mu(t_i) + \sum_{k=1}^g \xi_{i,k}\phi_k(t_i) + \epsilon_i$, where $\xi_{i,k}\sim \mathcal{N}(0,\lambda_k)$ and $\epsilon_i\sim \mathcal{N}\{0, \sigma^2(t_i)\}$ are drawn independently; again, each trajectory is censored so that all $w_i(t_{i,j})\in[-1,1]$. The trajectories for \textit{Cases 7} - \textit{10} are constructed in same manner as \textit{Cases 5} and \textit{6}. The FPC scores for \textit{Case 7} are simulated from a Gaussian mixture model with most of the density concentrated away from zero \citep{Yaoetal2005}. A Laplace distribution is used for \textit{Case 8} and a skew normal distribution is selected for \textit{Case 9}. Skewness parameters are chosen to exaggerate the skewness observed in the application data. Non-Gaussian measurement error in \textit{Case 10} is drawn from a shifted gamma distribution. All model parameters are selected to maintain $\E(\xi_{i,k})=0$ and $\text{Var}(\xi_{i,k})=\lambda_k$. Eigenfunctions, eigenvalues, and measurement error variance numbers match those from \textit{Case 6}, so direct comparisons of \textit{Case 6} to \textit{Cases 7} - \textit{10} may be made. Additional details are provided below.
\vspace{-.1cm}\begin{itemize}
\item[] (\textit{Case 1}) $\Sigma(T,T) = 0.5I$ and $\sigma^2(T)=0.05$.
\item[] (\textit{Case 2}) $\Sigma(T,T)$ has an AR(1) correlation structure with correlation parameter $\rho = 0.9$ and variance parameter $\psi = 0.5$. Again, $\sigma^2(T)=0.05$.
\item[] (\textit{Case 3}) $\Sigma(T,T)$ has a block diagonal structure. An AR(1) structure is enforced for all diagonal $g/3\times g/3$-dimensional blocks with variance $\psi = 0.5$. Let $\rho_1 = 0.5$ for the upper left block, $\rho_2=0.7$ for the middle, and $\rho = 0.9$ for the bottom right. The measurement error variance is $0.05$.
\item[] (\textit{Case 4}) $\Sigma(T,T)$ has the same structure as in \textit{Case 2}, except the measurement error is doubled in regions where $\mu(\cdot) < -0.5$ and halved in regions where $\mu(\cdot) > 0.5$ to resemble the estimated measurement error from the device used in our application \citep{Nagletal2021}.
\item[] (\textit{Case 5}) We construct $\Sigma(T,T)$ from eigenfunctions $\phi_1(t)= -\surd2\cos(\pi t)$, $\phi_2(t)= \surd 2\sin(\pi t)$, and $\{\phi_k(t)\}_{k=3}^g$ (for $k\geq 3$, the eigenfunctions are randomly generated and orthogonalized using QR factorization). Corresponding eigenvalues are of the form $\lambda_k = 1/(4k^2)$ and $\sigma^2(T)=0.05$.
\item[] (\textit{Case 6}) We construct $\Sigma(T,T)$ from the same eigenfunctions used in \textit{Case 5}. The corresponding eigenvalues are of the form $\lambda_k = 1/(4k^2)$ for $k=1,2,3,4$ and $\lambda_k = 1/(4k^3)$ for $k=5, \dots, g$, implying that 99\% the variability is captured from the first five FPC scores. The measurement error variance is the same as in \textit{Case 5}.
\item[] (\textit{Case 7}) Each $\xi_{i,k}$ is generated from either $N(-\sqrt{\lambda_k / 2}, \lambda_k / 2)$ or $N(\sqrt{\lambda_k / 2}, \lambda_k / 2)$ with equal probability \citep{Yaoetal2005}.
\item[] (\textit{Case 8}) Each $\xi_{i,k}$ is drawn from the Laplace distribution with the location parameter set to $0$ and the scale parameter set to $\sqrt{\lambda_k/2}$.
\item[] (\textit{Case 9}) Each $\xi_{i,k}$ is drawn from a skew normal distribution with location, shape, and scale parameters chosen such that $\E(\xi_{i,k})=0$, $Var(\xi_{i,k})=\lambda_k$. The skewness at each timepoint is set to match the maximum absolute skewness of the application data, meaning this simulation setting generates data more skewed than those from the application.
\item[] (\textit{Case 10}) The mean and covariance functions are identical to \textit{Case 6}, however the measurement error at each time index is generated from a gamma distribution with shape and rate parameters set to $1$ and $\sqrt{20}$, then shifted to have mean zero.
\end{itemize}
\vspace{-.1cm}On average, one third of the measurements are censored for all settings. Two GFLM settings are considered: the continuous response $Y_i= \sum\limits_{k\geq 1}\beta_k\xi_{i,k} + \omega_i$ with $\omega_i\overset{iid}{\sim}\mathcal{N}(0,0.1)$ and the identity link function, and the binary response, $Y_i= \ind\{\sum\limits_{k\geq 1}\beta_k\xi_{i,k} > 0\}$, with the logit link function. For both cases, we choose $\beta(t)=1$.

Table \ref{t:table1} shows the SSE for the mean and covariance estimators, averaged over 500 MC samples, for all three estimation procedures. The proposed mean and covariance estimators perform best for all cases. Table \ref{t:table2} shows the SSE for the measurement error variance estimator and the SNRs, which are defined by $\widehat{\text{SNR}}_n=\widehat{\sigma}_n^2(t)/\sigmasqWtildehat(t)$ at time $t$. The measurement error variance SSE is larger for $\widehat{\sigma}^{2}(\cdot)$ than for either $\widehat{\sigma}^{2,\N}(\cdot)$ or $\widehat{\sigma}^{2,\CE}(\cdot)$, however, this can be explained by the SNRs. All cases show favorable results for our proposed method based on the SNR, however both the naïve and PACE procedures underestimate the diagonal of $\SigmaWtilde (\cdot,\cdot)$, leading to smaller measurement error variance estimates, but inaccurate SNRs. When compared to \textit{Case 6}, simulations run under non-Gaussian FPC score and measurement error settings (\textit{Cases 7} - \textit{10}) produce similar results, suggesting that the procedure is robust to data generation procedures that violate modeling assumptions. Additionally, our method demonstrates better recovery of dominant eigenfunctions than the comparable procedures (see the Supplementary Material).


\begin{table}[H]
\vspace*{-6pt}
\centering
\def\~{\hphantom{0}}
\caption{SSE for $\widehat{\mu}_n(\cdot)$ and $\widehat{\Sigma}_n(\cdot,\cdot)$, averaged over 500 MC samples. For parameter $\theta$, $\widehat{\theta}_n$ is our proposed estimator, $\widehat{\theta}_n^{\N}$ is the naïve estimator, and $\widehat{\theta}_n^{\CE}$ is computed under the canonical PACE method settings. The MC standard error for each estimator is in parentheses.}\label{t:table1}
\begin{tabular*}{\columnwidth}{l@{\extracolsep{\fill}} 
c@{\extracolsep{\fill}}
c@{\extracolsep{\fill}}
c@{\extracolsep{\fill}}
c@{\extracolsep{\fill}}
c@{\extracolsep{\fill}}
c@{\extracolsep{\fill}}}
Case & $\widehat{\mu}_n(\cdot)$ & $\widehat{\mu}_n^{\N}(\cdot)$ & $\widehat{\mu}_n^{\CE}(\cdot)$ & $\widehat{\Sigma}_n(\cdot,\cdot)$ & $\widehat{\Sigma}_n^{\N}(\cdot,\cdot)$ & $\widehat{\Sigma}_n^{\CE}(\cdot,\cdot)$\\ 
\textit{Case 1 } & 0.11 (0.05) & 0.63 (0.03) & 0.68 (0.03) & 2.27 (0.23) & 2.49 (0.15) & 3.80 (0.13) \\
\textit{Case 2 } & 0.14 (0.06) & 0.62 (0.04) & 0.70 (0.04) & 1.59 (0.82) & 9.27 (0.32) & 8.20 (0.31) \\
\textit{Case 3 } & 0.14 (0.06) & 0.62 (0.04) & 0.68 (0.04) & 1.71 (0.58) & 4.19 (0.24) & 4.80 (0.23) \\
\textit{Case 4 } & 0.15 (0.06) & 0.63 (0.04) & 0.71 (0.04) & 1.57 (0.81) & 9.36 (0.31) & 8.21 (0.31) \\
\textit{Case 5 } & 0.12 (0.05) & 0.49 (0.04) & 0.55 (0.03) & 1.23 (0.60) & 4.30 (0.25) & 3.74 (0.25) \\
\textit{Case 6 } & 0.11 (0.05) & 0.44 (0.04) & 0.51 (0.03) & 1.06 (0.55) & 4.18 (0.24) & 3.43 (0.23) 
\end{tabular*}
\end{table}


\begin{table}[H]
\vspace*{-6pt}
\centering
\def\~{\hphantom{0}}
\caption{SSE for $\widehat{\sigma}_n^2(\cdot)$ and $\widehat{\text{SNR}}_n$, averaged over 500 MC samples. For parameter $\theta$, $\widehat{\theta}_n$ is our proposed estimator, $\widehat{\theta}_n^{\N}$ is the naïve estimator, and $\widehat{\theta}_n^{\CE}$ is computed under the canonical PACE method settings. The MC standard error for each estimator is in parentheses.}\label{t:table2}
\begin{tabular*}{\columnwidth}{l@{\extracolsep{\fill}} 
c@{\extracolsep{\fill}}
c@{\extracolsep{\fill}}
c@{\extracolsep{\fill}}
c@{\extracolsep{\fill}}
c@{\extracolsep{\fill}}
c@{\extracolsep{\fill}}}
Case & $\widehat{\sigma}_n^2(\cdot)$ & $\widehat{\sigma}_n^{2,\N}(\cdot)$ & $\widehat{\sigma}_n^{2,\CE}(\cdot)$ & $\widehat{\text{SNR}}_n$ & $\widehat{\text{SNR}}_n^{\N}$ & $\widehat{\text{SNR}}_n^{\CE}$ \\ 
\textit{Case 1 } & 0.96 (0.05) & 0.37 (0.03) & 0.57 (0.01) & 2.47 (0.06) & 5.35 (0.10) & 13.11 (0.06) \\
\textit{Case 2 } & 0.05 (0.02) & 0.01 (0.01) & 
0.00 (0.01) & 0.12 (0.03) & 0.30 (0.04) & 0.36 (0.03) \\
\textit{Case 3 } & 0.45 (0.04) & 0.09 (0.02) & 0.12 (0.01) & 1.25 (0.07) & 1.91 (0.07) & 3.89 (0.05) \\
\textit{Case 4 } & 0.05 (0.02) & 0.02 (0.01) & 0.01 (0.01) & 0.13 (0.03) & 0.31 (0.04) & 0.38 (0.03) \\
\textit{Case 5 } & 0.08 (0.02) & 0.02 (0.01) & 0.01 (0.00) & 0.30 (0.05) & 0.59 (0.06) & 1.12 (0.04) \\
\textit{Case 6 } & 0.03 (0.02) & 0.01 (0.01) & 0.00 (0.00) & 0.11 (0.04) & 0.26 (0.05) & 0.45 (0.03) 
\end{tabular*}
\end{table}


For \textit{Cases 5} - \textit{10}, we investigate trajectory prediction and FPC score recovery across all three methods. Table \ref{t:table3} shows the mean squared prediction error (MSPE). Addressing the censoring leads to better recovery of latent trajectories, even across timepoints that were not recorded. Even when the FPC scores were generated from a mixture distribution (\textit{Case 7}) or the measurement error was generated from a shifted gamma distribution (\textit{Case 10}), the prediction error is almost identical to that of the baseline setting (\textit{Case 6}). Additionally, the proposed FPC score estimator outperforms that of the naïve and PACE methods, especially when the distributional assumptions are not met (Table \ref{t:table4}) Robustness to non-Gaussian generative models may be partially attributed to the form of the score predictor. The predictor $\xi^*_{i,k}$ is defined as the conditional expectation of $\xiCE_{i,k}$, given the noisy censored trajectory and censoring indicators, and $\xiCE_{i,k}$ is the best linear unbiased predictor of $\xi_{i,k}$ regardless of Gaussian assumptions (Yao et al., 2005). Accurate estimation under such conditions is then dependent on how close the mean of the Truncated Normal distribution is to $\E \{\widetilde{W}(T)\mid W(T), \Delta(T)\}$, and how well the first and second moments of the distribution of $\widetilde{W}(T)$ are approximated when (incorrectly) assuming a Gaussian model.

\begin{table}[H]
\centering
\def\~{\hphantom{0}}
\caption{MSPE $= \tfrac{1}{n} \sum_{i=1}^n \sum_{j=1}^N \{Z_i(T_{i,j}) - \widehat{W}^K_i(T_{i,j})\}^2$, averaged over 500 MC samples. Results computed under the naïve and PACE estimation procedures are denoted by the superscripts 'N' and 'CE'.}\label{t:table3}
\begin{tabular*}{.45\linewidth}{l c c c}
Case & MSPE & MSPE$^{\text{N}}$ & MSPE$^{\CE}$ \\ 
\textit{Case 5} & 1.56 & 2.35 & 2.64 \\
\textit{Case 6} & 1.17 & 1.89 & 2.02 \\
\textit{Case 7} & 1.17 & 1.83 & 1.97 \\
\textit{Case 8} & 1.25 & 2.07 & 2.20 \\
\textit{Case 9} & 1.24 & 1.68 & 1.80 \\
\textit{Case 10} & 1.15 & 1.84 & 1.98 
\end{tabular*}
\end{table}

\begin{table}[H]
\centering
\def\~{\hphantom{0}}
\caption{SSE for $\widehat{\xi}_{n,i,k}$, $\sum_{i=1}^n \sum_{k=1}^g (\xi^*_{i,k}-\widehat{\xi}_{i,k,n})^2/g$, averaged over 500 MC samples. For parameter $\theta$, $\widehat{\theta}_n$ is our proposed estimator, $\widehat{\theta}_n^{\N}$ is the naïve estimator, and $\widehat{\theta}_n^{\CE}$ is computed under the canonical PACE method settings. The MC standard error for each estimator is in parentheses.}\label{t:table4}
\begin{tabular*}{.52\linewidth}{l c c c}
Case & $\widehat{\xi}_n$ & $\widehat{\xi}_n^{\text{N}}$ & $\widehat{\xi}_n^{\CE}$ \\ 
\textit{Case 5} & 0.86 (0.00) & 0.97 (0.01) & 9.49 (0.07) \\
\textit{Case 6} & 0.58 (0.00) & 0.70 (0.00) & 9.58 (0.07) \\
\textit{Case 7} & 0.59 (0.00) & 0.67 (0.00) & 9.65 (0.06)\\
\textit{Case 8} & 0.63 (0.01) & 0.83 (0.01) & 9.00 (0.08)\\
\textit{Case 9} & 0.62 (0.00) & 0.64 (0.00) & 9.99 (0.07)\\
\textit{Case 10} & 0.59 (0.00) & 0.70 (0.00) & 9.73 (0.07)
\end{tabular*}
\end{table}

Simulation results for the GFLM, including the functional coefficient and regression diagnostics, are computed for a range of fraction of variance explained settings and displayed in the Supplementary Material. For the continuous response case, the incorporation of censoring information results in more accurate functional coefficient estimates. The proposed method shows a slight improvement across regression diagnostics, however the three methods are comparable. This is potentially due to the symmetric construction of $\mu(\cdot)$ and the form of the GFLM; under the naïve approach, the use of observations censored at the upper and lower bounds may 'average out' the bias introduced from not accounting for censoring at either bound.

Finally, we compare our approach to the numeric approximation methods of Liu \& Houwing-Duistermaat (2022; 2023). We make the following adjustments to the mean and covariance estimation procedures so that they are comparable to our method: 1) the bandwidth selection methods detailed in Section 3 are used (as opposed to the oracle bandwidth used in the 2022 and 2023 papers) and 2) the covariance estimation method (Liu \& Houwing-Duistermaat, 2023) does not guarantee a PSD matrix, so we reconstruct the estimate after removing negative eigenvalues, if necessary.


We focus on the sparse simulation setting from Liu \& Houwing-Duistermaat (2023). For each trajectory, between 3 and 10 timepoints are sampled from a grid of 20 equally spaced points in the interval $[0,1]$, and trajectories are constructed using $\mu(t)=0$, $\Sigma(s,t)=4 \cos(4 \pi s) cos(4 \pi t)$, and $\sigma^2(t) = 1$. Originally, the timepoints were sampled from a grid of 200 equally spaced points, but we reduce that number to generate sufficient data compatible with a non-oracle bandwidth selection method. One hundred trajectories are simulated for each of the 100 MC samples and the trajectories are censored from below at 0. The mean function is estimated at times $t\in\{0, 1/19,2/19,\dots,1\}$ and the covariance function is estimated at all time pairs $(s,t)$ for $s,t\in\{0, 1/19, 2/19, \dots, 1\}$. When implementing the mean and covariance estimation methods from Liu \& Houwing-Duistermaat (2022; 2023), we use the OBS weighting scheme because it was reported to perform best for sparse data in the covariance setting and almost identically to the best estimator for the mean. The constant approximation estimator from Liu \& Houwing-Duistermaat (2022) resulted in an average $MSE_{\mu} = \tfrac{1}{20}\sum_{j=1}^{20}\{\mu(t_j) - \widehat{\mu}_n(t_j)\}^2$ of $0.18$ compared to $0.01$ from our procedure. Our covariance procedure is initialized under a set of potential starting matrices, and the bandwidth–starting matrix pair that maximized the pseudo-likelihood criterion from Section 3 is selected. The $MSE_{\Sigma} = \tfrac{1}{400}\sum_{j=1}^{20}\sum_{k=1}^{20}\{\Sigma(t_j,t_k) - \widehat{\Sigma}_n(t_j,t_k)\}^2$ is estimated under both methods and our covariance estimator achieved better results, yielding an average MSE of 0.71. In comparison, the average MSE from the adapted Liu \& Houwing-Duistermaat (2023) approach is 1.07.

\section{Application: Eating Disorder Classification}\label{sec5}
Individuals with type 1 diabetes (T1D) and poor glycemic control exhibit blood glucose readings that are both more variable and higher on average than those of individuals with T1D and better glycemic control. These concerns are typically exaggerated if the patient with T1D has an eating disorder (ED) diagnosis or symptomatology. Insufficient insulin dosing prevents the utilization of glucose for energy and promotes the breakdown of fat as an alternative energy source; consequently, insulin omission alone, or following an overeating episode, may be a sign of disordered behavior in patients with T1D that results in elevated post-prandial blood glucose levels for an abnormally long period of time. Coupled with appropriate diagnostic instruments, glucose trajectory characteristics may be used to distinguish between individuals with and without EDs. Continuous glucose monitors (CGMs) estimate blood glucose levels from interstitial fluids, providing a trajectory of glucose recordings across a discretized grid of times, which can be used to inform diabetes management decisions. However, the accompanying CGM software truncates readings to ``High" if the recording is $\geq 400$ mg/dL and ``Low" if it is $\leq 40$ mg/dL (from a clinical perspective, censoring is reasonable because all values outside of the range are dangerous to the patient). Therefore, instead of observing an individual's true blood glucose trajectory, we record noisy, and possibly censored, readings in five minute increments. The inclusion of censored points is physiologically relevant because a complete case approach would artificially exclude trajectories from individuals with severe ED symptoms.

\begin{figure}[H]
  \centering
  {\includegraphics[width=0.5\textwidth]{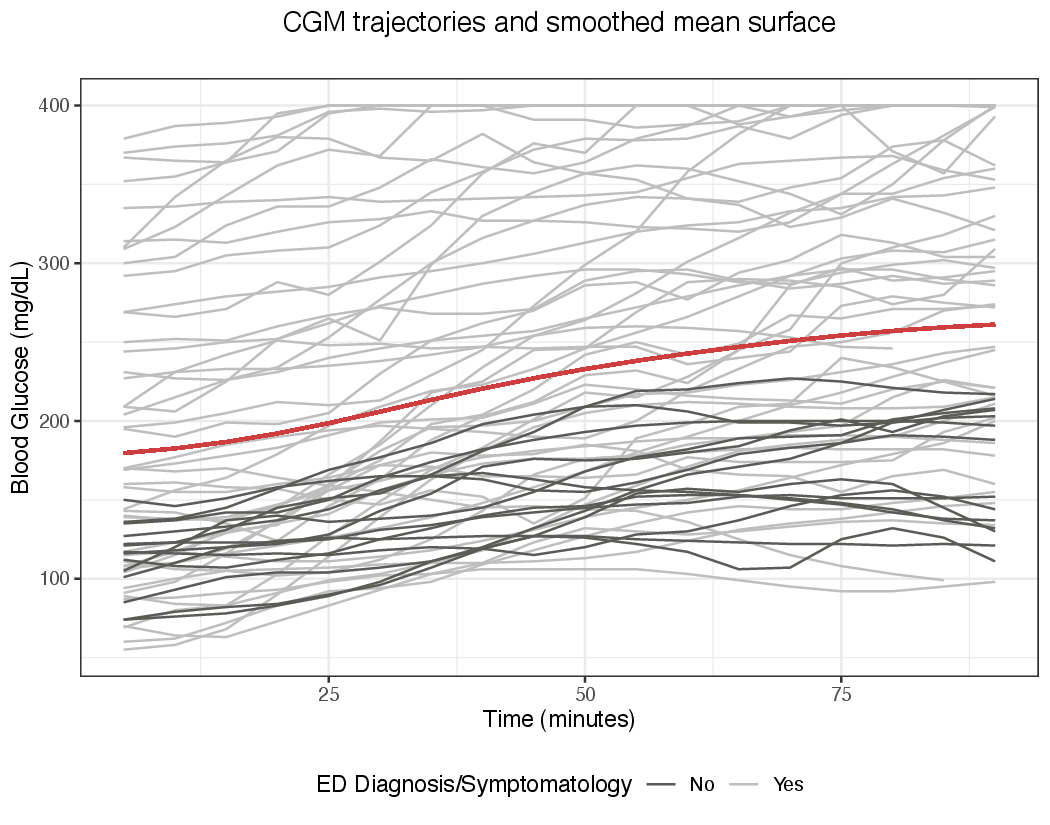}}
  \hspace*{\fill}
  {\includegraphics[width=0.47\textwidth]{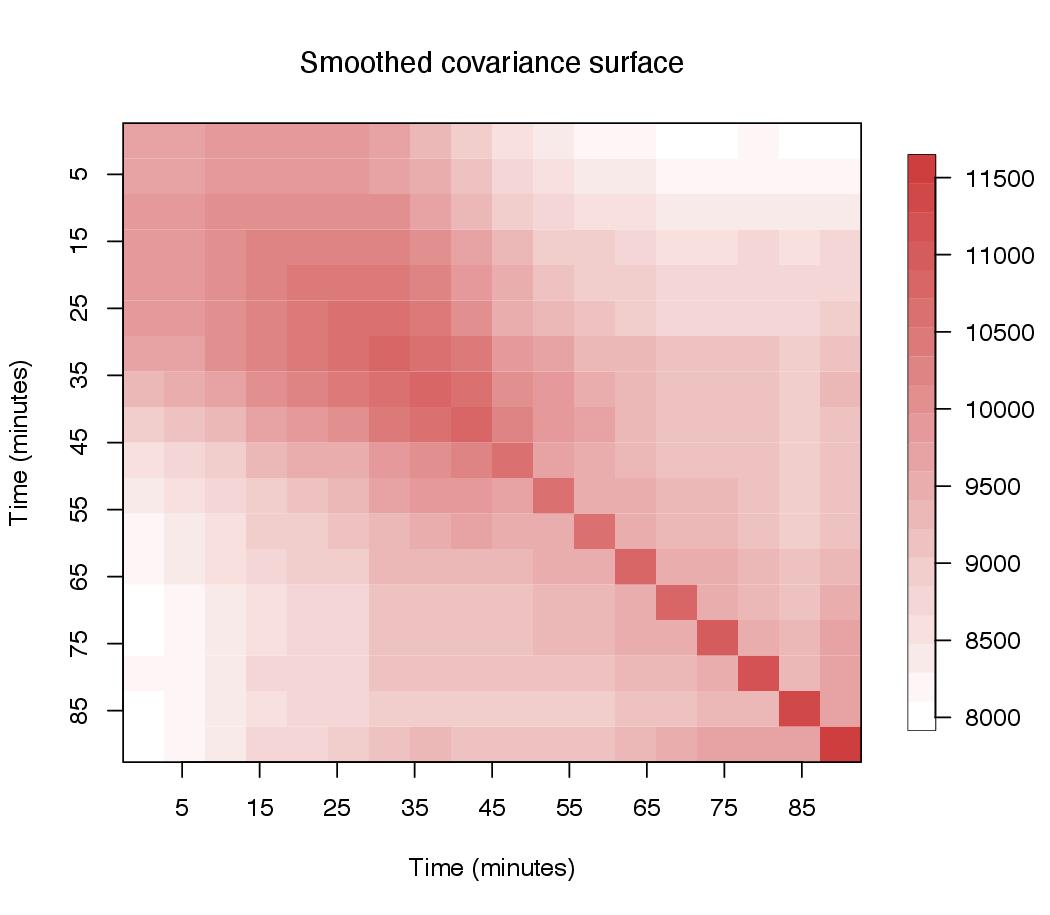}}
  \caption{Left: Blood glucose trajectories (grey) with the estimated mean function (red) overlayed. Right: Estimated covariance surface.}\label{fig:muhatED}
\end{figure}

We analyze the following subset of data from the Eating Disorders in Type 1 Diabetes: Mechanisms of Comorbidity clinical study (R01 DK089329, PI: Merwin): 90 minutes of post-meal CGM recordings, baseline characteristics, and ED status for $n=59$ patients ($45$ of whom have a diagnosed ED). Data for each study participant are of a form similar to those discussed in Section \ref{sec3}, and 10 of the 59 trajectories are subject to censoring (approximately 7\% of the observed points). Less than 1\% of the data are missing; all missingness is attributed to a brief disconnection between the CGM and accompanying software. Observation times represent the time difference between a CGM recording and the start of the meal, and are re-scaled so that all differences fall in the interval $[0,1]$. Baseline covariates include HbA1C, age of T1D diagnosis, and age. The outcome of interest is the ED indicator (1 if patient $i$ has an ED diagnosis or symptomatology, and 0 otherwise). Figure \ref{fig:muhatED} shows the glucose recordings with the smoothed mean estimate overlayed in red (left), as well as the covariance surface estimate (right). On average, glucose levels begin to spike almost immediately post meal, then level off after approximately 75 minutes. As expected, the T1D patients without EDs run lower on average than their ED counterparts. We train the binary response GFLM using 5-fold cross-validation, and the average accuracy across test sets is 0.83. Additionally, we consider 1) a regression on the average of the CGM trajectory and 2) omitting the CGM data entirely. In comparison to the GFLM, both models display poor accuracy, AIC, and residual deviance (see the Supplementary Material for additional results).

\section{Discussion}\label{sec6}
{\color{editColor}We presented a procedure to estimate FPC scores, centered around the accurate recovery of smoothed mean and covariance functions, that is appropriate for censored and non-censored data, with measurement error. 
The proposed approach appeared to be robust to violations of normality, however, relaxing the parametric assumptions on the censored data is of interest. One possible approach might be to estimate the conditional distribution of the censored trajectory, given other available covariates, using a framework similar to that of Kong and Nan (2016). However, temporal correlation would have to be incorporated, and practically, the method might be difficult to implement with limited covariate data.}

While irregularly spaced and sparsely observed data were considered, extensions to 1) data requiring trajectory alignment and 2) settings for which units have repeated functional measurements, so-called second-generation functional data \citep{KonerStaicu2023}, were not addressed. The former is particularly relevant if there are external events that may alter the functional process at any period in the observation window. If event times are random, the resulting trajectories may be misaligned in time. A variety of time warping techniques have been proposed to address this issue for the non-censored case \citep{Wangetal2016}. Extensions of this methodology to the censored data case are particularly applicable to longer trajectory data, such as daily blood glucose recordings for which meal times are not aligned.

\section*{Supplementary Material}
\label{SM}
The Supplementary Material include the referenced results on partitioned multivariate normal vectors, the covariance optimization algorithm, the proof of Theorem 1, an example of non-Gaussian measurement error, and additional simulation and application results.

\bibliographystyle{biometrika}
\bibliography{paper-ref}

\end{document}